\documentclass[aps,pra,reprint,superscriptaddress]{revtex4-2}

\usepackage{amsmath}
\usepackage{graphicx}
\usepackage{multirow}
\usepackage{booktabs}
\usepackage[colorlinks,
linkcolor=blue,
anchorcolor=blue,
citecolor=blue,
urlcolor=blue	
]{hyperref}
\usepackage{bm}
\usepackage{dcolumn}
\usepackage{siunitx}
\usepackage{xcolor}

\newcolumntype{d}[1]{D{.}{.}{#1}}

\begin{document}

\title{Precise determination of the $\mathbf{2s^{2}2p^{5}-2s\,2p^{6}}$ transition energy in fluorine-like nickel utilizing a low-lying dielectronic resonance}

\author{S. X. Wang}
\affiliation{Department of Modern Physics, University of Science and Technology of China, Hefei, Anhui 230026, China}

\author{Z. K. Huang}
\affiliation{Institute of Modern Physics, Chinese Academy of Sciences, 730000, Lanzhou, China}

\author{W. Q. Wen}\email[]{wenweiqiang@impcas.ac.cn}
\affiliation{Institute of Modern Physics, Chinese Academy of Sciences, 730000, Lanzhou, China}
\affiliation{University of Chinese Academy of Sciences, 100049, Beijing, China}

\author{W. L. Ma}
\affiliation{Department of Modern Physics, University of Science and Technology of China, Hefei, Anhui 230026, China}

\author{H. B. Wang}
\affiliation{Institute of Modern Physics, Chinese Academy of Sciences, 730000, Lanzhou, China}

\author{S. Schippers}
\affiliation{I. Physikalisches Institut, Justus-Liebig-Universit\"at Gie{\ss}en, 35392 Giessen, Germany}
\affiliation{Helmholtz Forschungsakademie Hessen f\"ur FAIR (HFHF), Campus Giessen, 35392 Giessen, Germany}

\author{Z. W. Wu}
\affiliation{Key Laboratory of Atomic and Molecular Physics $\&$ Functional Materials of Gansu Province, College of Physics and Electronic Engineering, Northwest Normal University, Lanzhou 730070, China}

\author{Y. S. Kozhedub}\author{M. Y. Kaygorodov}
\affiliation{Department of Physics, St. Petersburg State University, Universitetskaya 7/9, 199034 St. Petersburg, Russia}

\author{A. V. Volotka}
\affiliation{School of Physics and Engineering, ITMO University, Kronverkskiy prospekt 49, 197101 St. Petersburg, Russia}

\author{K. Wang}
\affiliation{Hebei Key Lab of Optic-Electronic Information and Materials, The College of Physics Science and Technology, Hebei University, Baoding 071002, China}

\author{C. Y. Zhang}\author{C. Y. Chen}
\affiliation{Shanghai EBIT Laboratory, Institute of Modern Physics, Fudan University, Shanghai 200433, China}

\author{C. Liu}
\affiliation{Department of Modern Physics, University of Science and Technology of China, Hefei, Anhui 230026, China}

\author{H. K. Huang}
\affiliation{Institute of Modern Physics, Chinese Academy of Sciences, 730000, Lanzhou, China}
\affiliation{University of Chinese Academy of Sciences, 100049, Beijing, China}

\author{L. Shao}
\affiliation{Institute of Modern Physics, Chinese Academy of Sciences, 730000, Lanzhou, China}

\author{L. J. Mao}
\affiliation{Institute of Modern Physics, Chinese Academy of Sciences, 730000, Lanzhou, China}
\affiliation{University of Chinese Academy of Sciences, 100049, Beijing, China}

\author{X. M. Ma}\author{J. Li}\author{M. T. Tang}\author{K. M. Yan}\author{Y. B. Zhou}
\affiliation{Institute of Modern Physics, Chinese Academy of Sciences, 730000, Lanzhou, China}

\author{Y. J. Yuan}\author{J. C. Yang}
\affiliation{Institute of Modern Physics, Chinese Academy of Sciences, 730000, Lanzhou, China}
\affiliation{University of Chinese Academy of Sciences, 100049, Beijing, China}

\author{S. F. Zhang}\author{X. Ma}
\email[]{x.ma@impcas.ac.cn}
\affiliation{Institute of Modern Physics, Chinese Academy of Sciences, 730000, Lanzhou, China}
\affiliation{University of Chinese Academy of Sciences, 100049, Beijing, China}

\author{L. F. Zhu}
\email[]{lfzhu@ustc.edu.cn}
\affiliation{Department of Modern Physics, University of Science and Technology of China, Hefei, Anhui 230026, China}

\date{\today}

\begin{abstract}
High precision spectroscopy of the low-lying dielectronic resonances in fluorine-like Ni$^{19+}$ ions was studied 
by employing the electron-ion merged-beams method at the heavy-ion storage ring CSRm. The measured 
dielectronic-recombination (DR) resonances are identified by comparison  with  relativistic calculations utilizing the 
flexible atomic code (FAC). The lowest-energy resonance at about 86~meV is due to DR via 
$(2s2p^{6}[^{2}S_{1/2}]6s)_{J=1}$ intermediate state. The  position of this resonance could be determined within an 
experimental uncertainty of as low as $\pm$4 meV. The binding energy of the 6$s$ Rydberg electron in the resonance 
state was calculated using two different approaches, the Multi-Configurational Dirac-Hartree-Fock (MCDHF) method and 
the Stabilization Method (SM). The sum of the experimental  $(2s2p^{6}[^{2}S_{1/2}]6s)_{J=1}$ resonance energy and the 
theoretical $6s$ binding energies from the MCDHF and SM calculations, yields the following values for the  
$2s^{2}2p^{5}\;^{2}P_{3/2} \to 2s2p^{6}\;^{2}S_{1/2}$ transition energy149.056(4)$_{\rm exp}$(20)$_{\rm theo}$ and 
149.032(4)$_{\rm exp}$(6)$_{\rm theo}$, respectively. The theoretical calculations reveal that second-order QED and 
third-order correlation effects contribute by together about 0.1~eV to the total transition energy and can, thus, be 
assessed by the present precision DR spectroscopic measurement.
\end{abstract}

\maketitle

\section{Introduction} \label{sec:intro}

Atomic energy levels of highly charged ions (HCIs) are ideal systems for testing the quantum electrodynamics (QED) and 
relativistic effects \cite{Safronova2018,Indelicato2019}. Electron-beam ion traps 
\cite{Beiersdorfer2009a,Beiersdorfer2010,Kozlov2018} and heavy-ion storage rings 
\cite{Grieser2012,Lestinsky2016,Steck2020} offer unique opportunities for precision studies with HCI. In particular, 
heavy-ion storage rings equipped with an electron cooler serve as ideal platforms for  electron-ion merged-beams 
experiments. This technique has been intensively  employed at  the TSR at MPIK in Heidelberg 
\cite{Wolf2006c,Schippers2015}, the CRYRING at MSL in Stockholm \cite{Schuch2007a} (in 2013 relocated to GSI in 
Darmstadt \cite{Lestinsky2016}), and the ESR at GSI  \cite{Brandau2013,Brandau2015}.  More recently, the experimental 
approach was also implemented at  the ion-storage rings HIRFL-CSRm and CSRe at the Institute of Modern Physics (IMP), 
Chinese Academy of Sciences, and delivered already results on electron-ion recombination of a number of ion species 
\cite{Huang2015,Huang2018,Wang2019}. Here, it is used for a precise determination of the $2s^22p^5\;^2P_{3/2}\to 
2s\,2p^6\;^2S_{1/2}$ transition energy in F-like Ni$^{19+}$ ions.

\begin{figure}[t]
	\centering
	\includegraphics[width=1\linewidth]{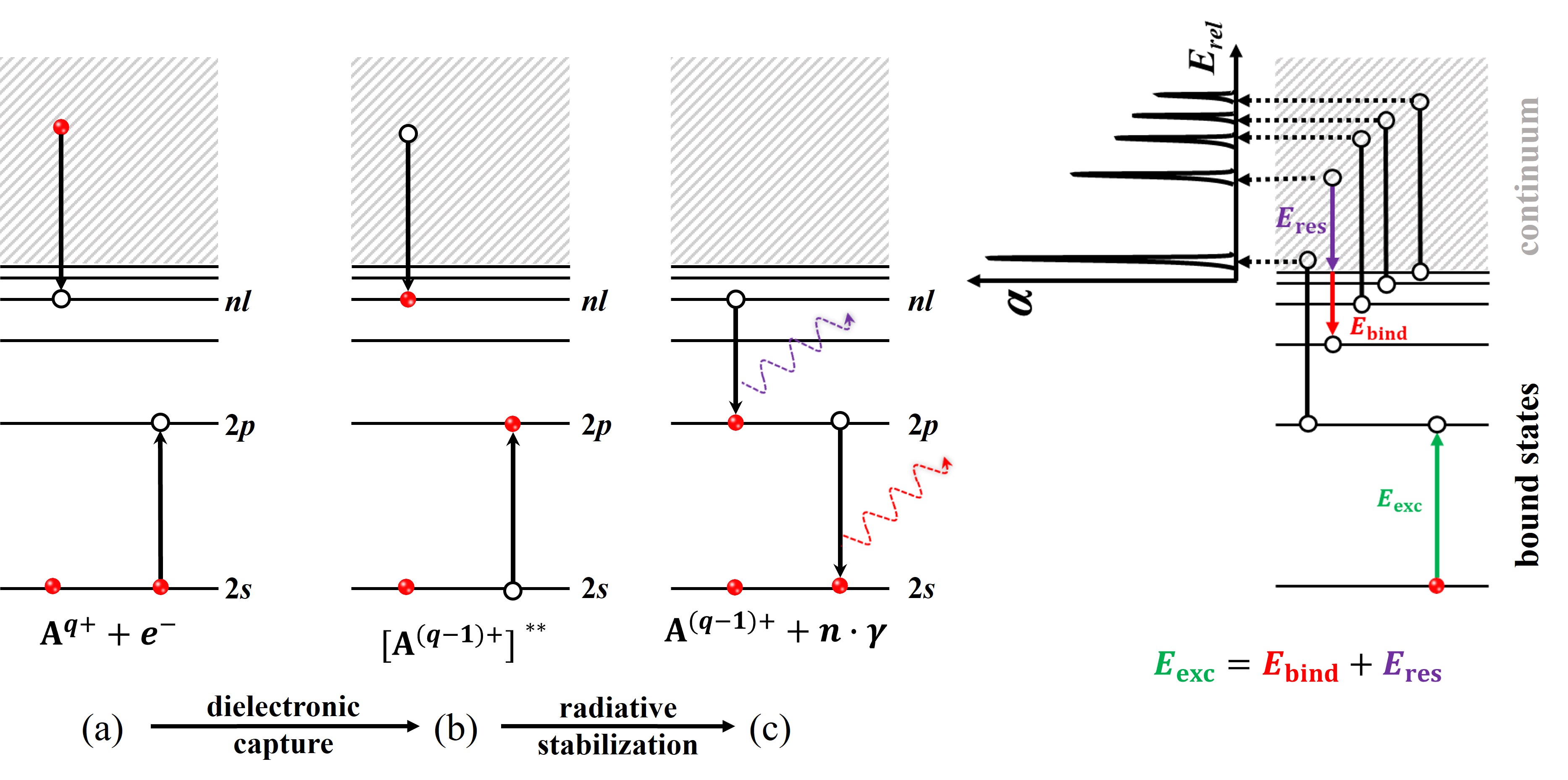}
	\caption{	\label{fig:DR}Left: Schematic diagram of the DR process. The step (a) $ \to $ (b) represents a 
	dielectronic capture: a free electron is resonantly captured into a Rydberg shell with simultaneous excitation of 
	an inner electron. The step (b) $ \to $ (c) depicts the subsequent radiative stabilization: the multiply excited 
	intermediate state stabilizes via photon emission. Right: By energy conversation  the core excitation energy 
	$E_\mathrm{exc}$ equals the sum of the Rydberg electron's binding energy $E_\mathrm{bind}$ and the DR resonance 
	energy $E_\mathrm{res}$.  The $\alpha$ vs. $E_\mathrm{rel}$ curve represents a made-up DR spectrum that illustrates 
	the principle of how a merged-beams DR rate coefficient $\alpha$ results from a scan  of the electron-ion collision 
	energy $E_\mathrm{rel}$ in the electron-ion center-of-mass frame.}
\end{figure}

The traditional approach to atomic precision spectroscopy of HCI is x-ray spectroscopy, which has been widely used for testing QED predictions concerning the atomic structure of HCI (see, e.g., \cite{Stoehlker1993,Crespo1996a,Beiersdorfer1998a,Stoehlker2000}). In a joint experimental and theoretical effort,  \citet{Lindroth2001} demonstrated that QED contributions to atomic transition energies can also be tested by electron-ion collision spectroscopy  at a heavy-ion storage ring,  where an electron-ion merged-beams arrangement  is employed for measuring low-energy dielectronic  recombination (DR) resonances with high experimental resolving power.  Moreover, electron-ion collision spectroscopy provides access to non-dipole transitions  which usually cannot easily be studied by optical spectroscopy \cite{Bernhardt2015a}.

In the DR process, a free electron is resonantly captured into a Rydberg state with simultaneous excitation of an inner electron (dielectronic capture, Fig.~\ref{fig:DR}). If the thus formed intermediate multiply excited state decays radiatively the DR process will be completed (radiative stabilization). The measured DR resonance energies $E_\mathrm{res}$ correspond to  the  energies of the associated doubly-excited states with respect to the first ionization energy of the recombining ion's initial level. As depicted in Fig.~\ref{fig:DR}, the core-excitation energy $E_\mathrm{exc}$ of the recombining ion can be obtained as
\begin{equation}\label{eq:energy}
	E_{\text{exc}}=E_{\text{res}}+E_{\mathrm{bind}},
\end{equation}
where $E_{\mathrm{bind}}$ is the absolute value of the Rydberg electron binding energy. For sufficiently high 
principal quantum numbers $n$ of the Rydberg electron the latter quantity can be calculated to a high accuracy, e.g., 
in the framework of relativistic many-body perturbation theory (RMBPT) \cite{Lindroth2001,Lestinsky2008a}. In 
principle, if several resonances of a Rydberg series are measured, $E_\mathrm{bind}$ can be extrapolated to zero by 
extrapolating $n\to\infty$ such that no separate calculation of $E_\mathrm{bind}$ is required 
\cite{Brandau2003b,Bernhardt2015}. In the present work, however, we employ the joint experimental and theoretical 
approach similar to what was used in earlier studies \cite{Lindroth2001,Madzunkov2002,Kieslich2004a,Lestinsky2008a}.

As a consequence of the merged-beams kinematics, the experimental resolving power of electron-ion merged-beams experiments is highest at the lowest energies (see, e.g., \cite{Mueller1999a}). At the same time, the uncertainty of the experimental electron-ion collision energy scale is lowest at zero collision energy (see, e.g., \cite{Bernhardt2015}).  Therefore, work on precision electron-ion collision spectroscopy has concentrated on measuring low-energy DR resonances.

In Ref.~\cite{Lindroth2001}, the $4p_{1/2} - 4s_{1/2}$ transition energy in copper-like Pb$^{53+}$  was determined to $E_\mathrm{exc}=118.101\pm0.001$~eV from measured DR resonances at $E_\mathrm{res} \lesssim 40$ meV  with an uncertainty of only $\pm$1~meV, corresponding to an accuracy of 8.5~ppm. In a follow-up study, \citet{Madzunkov2002} investigated the  $2p_{1/2} - 2s_{1/2}$ transition in Li-like Kr$^{33+}$. An uncertainty of $\pm$8~meV  was obtained for the transition energy. This comparatively large value was due to the fact that the lowest-energy DR resonances of this ion appear only  above about 5~eV. At the TSR, the $2p_{3/2} - 2s_{1/2}$ transition energy  in Li-like Sc$^{18+}$ could be determined to within $\pm$2~meV \cite{Kieslich2004a} from DR resonances in the electron-ion collision energy range 30--70~meV. Later, \citet{Lestinsky2008a}  reduced this uncertainty by more than an order of magnitude by using an internally cold electron beam from a liquid-nitrogen cooled photocathode \cite{Orlov2005a}. Their result probes few-body effects on radiative corrections on the 1\% level. Moreover, the experimentally achieved resolving power allowed for the observation of the hyperfine splitting of the  $1s^2\,2s\;^2S_{1/2}$ ground level.

Here, we present a precision-spectroscopy measurement with a more complex system than the previously studied Li-like ions, i.e., fluorine-like $^{58}$Ni$^{19+}$. As already reported in Ref.~\cite{Wang2019}, which focuses on providing atomic data for applications in  astrophysics, we have measured the merged-beams rate coefficient for DR of $^{58}$Ni$^{19+}$ ions in the energy of 0--160~eV, which included all resonances associated with $2s\to 2p$ core excitations ($\Delta N=0$ DR):
\begin{equation}
	\begin{aligned}
		&{^{58}\rm{Ni}^{19+}}(2s^{2}2p^{5}[^{2}P_{3/2}])+e^{-} \\
		&\rightarrow
		\begin{cases}
			&{^{58}\rm{Ni}^{19+}}(2s^{2}2p^{5}[^{2}P_{1/2}]nl)^{**} \rightarrow {^{58}\rm{Ni}^{18+}}+\gamma, \\
			&{^{58}\rm{Ni}^{19+}}(2s\,2p^{6}[^{2}S_{1/2}]nl)^{**} \rightarrow {^{58}\rm{Ni}^{18+}}+\gamma.
		 \end{cases}
	\end{aligned}
\end{equation}
The lowest-energy $(2s\,2p^{6}[^{2}S_{1/2}]\,6s)_{J=1}$ resonance occurs at $\sim$86~meV.  The currently recommended 
value for the associated $2s^{2}2p^{5}\;^{2}P_{3/2} - 2s\,2p^{6}\;^{2}S_{1/2}$ core-transition energy  is 
$149.05\pm0.12$~eV \cite{Kramida2021}. The results of the above described previous electron-ion collision-spectroscopic 
works suggest that the uncertainty of this value can be much reduced to a few meV, provided the binding energy of the 
6$s$ Rydberg electron can be evaluated to an even better accuracy. This would offer an opportunity for a sensitive test 
of second-order QED contributions to electron binding energies in fluorine-like nickel. We mention that previous work 
on the  $2s^{2}2p^{5}\;^{2}P_{3/2} - 2s\,2p^{6}\;^{2}S_{1/2}$ transition in Ni$^{19+}$ has been carried out using 
various experimental \cite{Doschek1974,Breton1979,Sugar1992} and theoretical 
\cite{Gu2005a,Joensson2013c,Nandy2014,Si2016,Fontes2017,Celik2020} approaches, partly addressing atomic-data needs in 
fusion-plasma physics and astrophysics.

The present paper is organized as follow. The experimental procedure is presented in Sec.~\ref{sec:exp} with a detailed discussion of the data-reduction procedures and the error analysis. In Sec.~\ref{sec:theory}, we present a general description of the theoretical treatment. The experimental results are then discussed in Sec.~\ref{sec:results}. Finally, Sec.~\ref{sec:sum} provides a conclusive summary.

\section{Experiment} \label{sec:exp}

\subsection{Measurement procedure}

The experiment was performed by employing the electron-ion merged-beams technique at the heavy-ion storage ring CSRm at 
the 
Institute of Modern Physics in Lanzhou, China. Several DR measurements related to astrophysical and plasma applications 
have been carried out successfully at the CSRm \cite{Huang2015,Huang2018,Wang2019} since the calibration experiment 
with lithium-like Ar$^{15+}$ in 2015 \cite{Huang2015}. Recombination rate-coefficients of fluorine-like nickel have 
already been published previously in Ref.~\cite{Wang2019}, which also contains a detailed description of the 
experimental setup and procedures. Here we  focus on the precise evaluation of the resonance energies of the 
lowest-energy  DR resonances.

In the present measurement, the $^{58}$Ni$^{19+}$ ion beam from a superconducting electron cyclotron resonance ion 
source \cite{Zhao2017} was accelerated by a sector focused cyclotron and then injected into the storage ring at an 
energy of 6.15~MeV/u. The stored ion beam reached a maximum current of 80~$\mu$A after the injection pulses, 
corresponding to $3.7\times10^{8}$ stored ions. The circulating ion beam passed millions of times per second through 
the 4~m long electron-ion interaction region. In the electron cooler, the electron beam is magnetically guided  to 
prevent it from diverging. The magnetic fields at the cathode and the cooler section were 125~mT and 39~mT, 
respectively. Thereby, the transverse electron beam energy spread is reduced by adiabatically passing from the higher 
magnetic field in the cathode to the lower magnetic field in the cooler section \cite{Danared1993b}. In the 
electron-ion interaction region, the expanded electron beam had a diameter of  62 mm and a particle density of 
7.1$\times10^{6}$ cm$^{-3}$.  By electron-beam profile measurements \cite{Bocharov2004} we verified that a uniform beam 
density distribution was achieved in the present measurements. The longitudinal electron energy-spread was largely 
reduced by accelerating the electrons to  the cooling energy, where the electrons move as fast as the stored ions.

The electron velocity-distribution can be characterized by the transverse (with respect to the electron-beam direction) and longitudinal temperatures $T_{\parallel} $ and $T_\perp$ \cite{Kilgus1992}:
\begin{equation}\label{eq:vdist}
	\begin{aligned}
		f(\vec{v},v_d) =& (\frac{m_{e}}{2\pi kT_{\parallel}})^{1/2} \exp \left[ -\frac{m_{e}(v_{\parallel}-v_d)^{2}}{2kT_{\parallel}} \right]\\
		& \times \frac{m_{e}}{2\pi kT_{\perp}}\exp\left( -\frac{m_{e}v_{\perp}^{2}}{2kT_{\perp}} \right).
	\end{aligned}
\end{equation}
The distribution is termed \emph{flattened Maxwellian} due to the fact that $T_{\parallel} \ll T_{\perp}$. Its 
anisotropy leads to asymmetric DR resonance line-shapes as discussed in more detail below. In Eq.~(\ref{eq:vdist}), 
$v_d$ corresponds to the energy detuning applied to the electron beam. $v_{\parallel}$ and $v_{\perp}$ are the 
longitudinal and perpendicular components of the electron velocity $\vec{v}$.
The electron-energy spread  that results from Eq.~(\ref{eq:vdist}) can be calculated as \cite{Mueller1999a}
\begin{equation}\label{eq2}
	\Delta E_\mathrm{rel} = \sqrt{(\ln2 \cdot k_{\text{B}}T_{\perp})^{2}+16\ln2 \cdot k_{\text{B}}T_{\parallel} \cdot E_\mathrm{rel}},
\end{equation}
where $E_\mathrm{rel}$ denotes the electron-ion collision energy  in the electron-ion center of mass frame. It is evident that the lowest-energy resonances can be measured with the highest resolving power.

\begin{figure}[t]
	\centering
	\includegraphics[width=\linewidth]{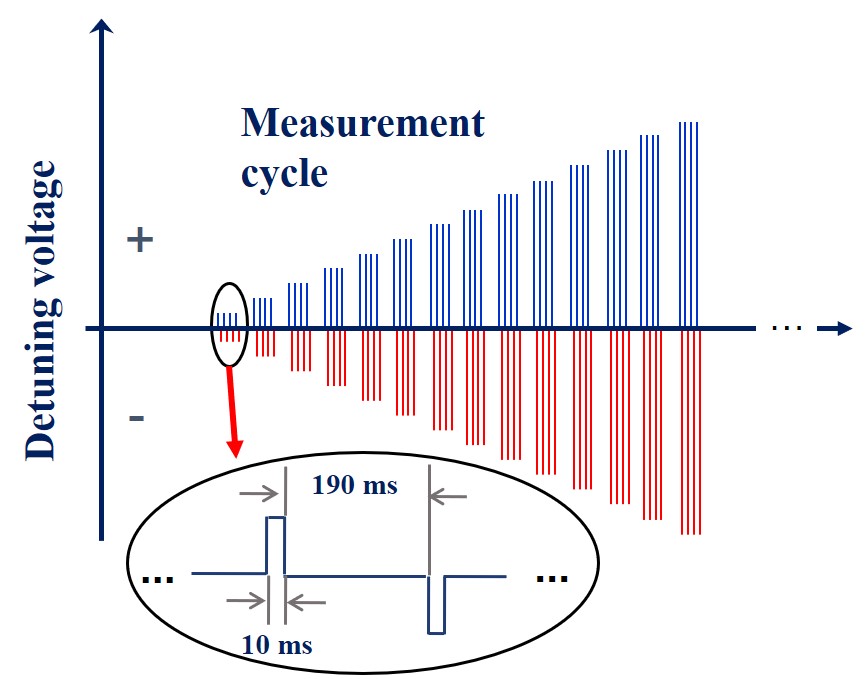}
	\caption{\label{fig:ramp}Illustration of the timing sequence of detuning voltages in the present measurement. The 
	voltage was incremented volt by volt only after each injection, i.e., we only measured  recombined ions at a single 
	detuning voltage during each injection-measurement cycle to collect a statistical significant number of counts.}
\end{figure}

\begin{figure}[t]
	\centering
	\includegraphics[width=1\linewidth]{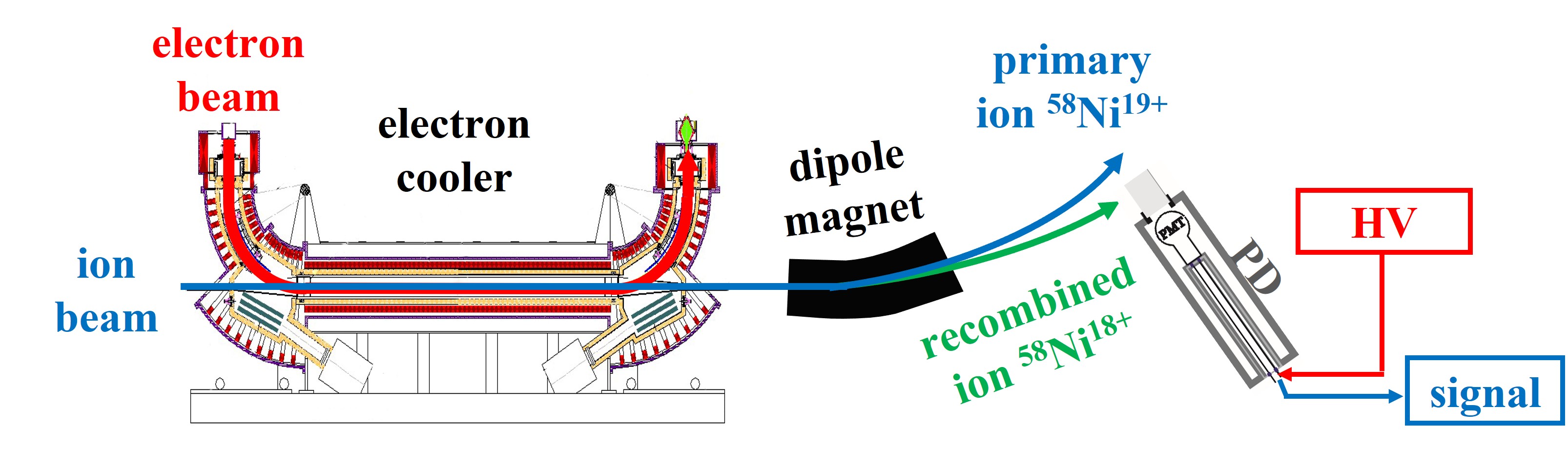}
	\caption{	\label{fig:setup}Sketch of the experimental arrangement at the CSRm electron cooler. The scintillation particle-detector was appropriately placed behind the dipole magnet downstream of the electron cooler for stopping and counting the recombined ions.}
\end{figure}

The injected ion beam was cooled by the Coulomb interaction with the cold electrons inside the electron cooler 
\cite{Poth1990}. At cooling, the electrons and the ions shared the same average velocity in the laboratory frame, 
corresponding to zero electron-ion collision energy in the center-of-mass-frame. The cooling phase at the beginning of  
each injection-measurement cycle lasted for 2~s. In the ensuing electron-ion recombination measurements, the cold 
electrons acted as an electron target.  Nonzero electron-ion collision  energies were realized by fast detuning the 
electron beam energy away from the cooling energy.  The electron beam energy was controlled by a especially  designed 
detuning system, which is capable of  switching quickly between the cooling voltage and positive or negative detuning 
voltages as illustrated in Fig.~\ref{fig:ramp}. After every 10~ms of detuning, the ion beam was cooled again for 190~ms 
to keep the beam quality. The recombined ions with a changed charge state lowered by one unit were separated from the 
primary ion beam in the dipole magnet downstream from the electron cooler and detected by a movable scintillation 
particle-detector (YAP: Ce+PMT) with nearly 100\% efficiency \cite{Wen2013}. A sketch of the electron cooler and the 
particle detector is presented in Fig.~\ref{fig:setup}. The ion current and the revolution frequency were monitored by 
a DC current transformer and a Schottky spectrum-analyzer \cite{Wu2013}, respectively.

\subsection{Data reduction}

The absolute recombination rate coefficient $\alpha(E)$ at the electron-ion collision energy $E_\mathrm{rel}$ was obtained by normalizing the recorded count rates $R(E_\mathrm{rel})$ to the measured electron and ion currents \cite{Bernhardt2011a}:
\begin{equation}\label{eq:alpha}
	\alpha(E_\mathrm{rel}) = \frac{R(E_\mathrm{rel})}{N_{i}n_{e}(1-\beta_{e}\beta_{i})}\frac{C}{L},
\end{equation}
where $N_{i}$ is the number of the stored ions and $n_e$ is the electron density, which is virtually  independent of 
the electron energy over the narrow energy range considered in this work. The quantities $C=161.0$ m and $L=4.0$ m are 
the circumference of the ring and the effective interaction length. $\beta_{e}$ and $\beta_{i}$ are the velocity 
factors of the electron and ion beams, respectively. The electron-ion collision energy in the center-of-mass-frame has 
been calculated as \cite{Schippers2000b}
\begin{equation}\label{eq:E}
	E_\mathrm{rel}  = m_{i}c^{2}(1+\mu) \cdot
	\left[
	\sqrt{1+\frac{2\mu}{(1+\mu)^{2}}(G-1)}-1
	\right] ,
\end{equation}
with the electron-ion mass-ratio $\mu=m_{e}/m_{i}$,
\begin{equation}\label{eq:G}
	G = \gamma_{e}\gamma_{i} - \sqrt{(\gamma_{e}^{2}-1)(\gamma_{i}^{2}-1)}\cos\theta,
\end{equation}
$\gamma_{e,i}=(1-\beta_{e,i}^2)^{-1/2}$, and the angle $\theta$ between the two beams in the laboratory frame.

\begin{figure}[t]
	\centering
	\includegraphics[width=1\linewidth]{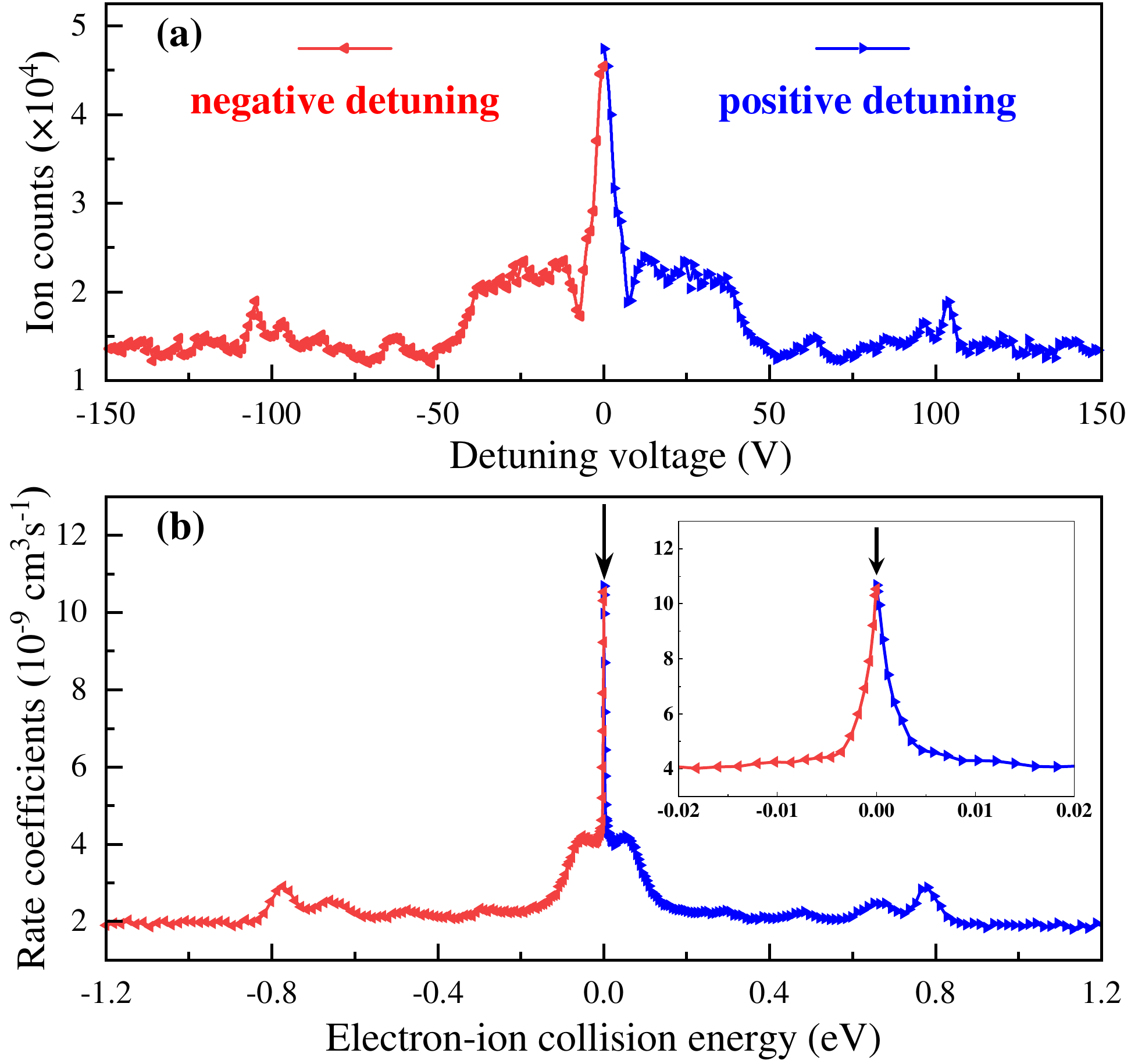}
	\caption{\label{fig:data}(a) The original recorded recombined ion counts for detuning voltages up to $\pm$150~V.	(b) The derived rate coefficients for the corresponding electron-ion collision energies ranging from  $-1.2$~eV to  $+1.2$~eV. The negative sign refers to  collision energies that result from negative detuning voltages. The inset highlights the recombination rate-coefficient near zero electron-ion collision energy.}
\end{figure}

\begin{figure}[t]
	\centering
	\includegraphics[width=1\linewidth]{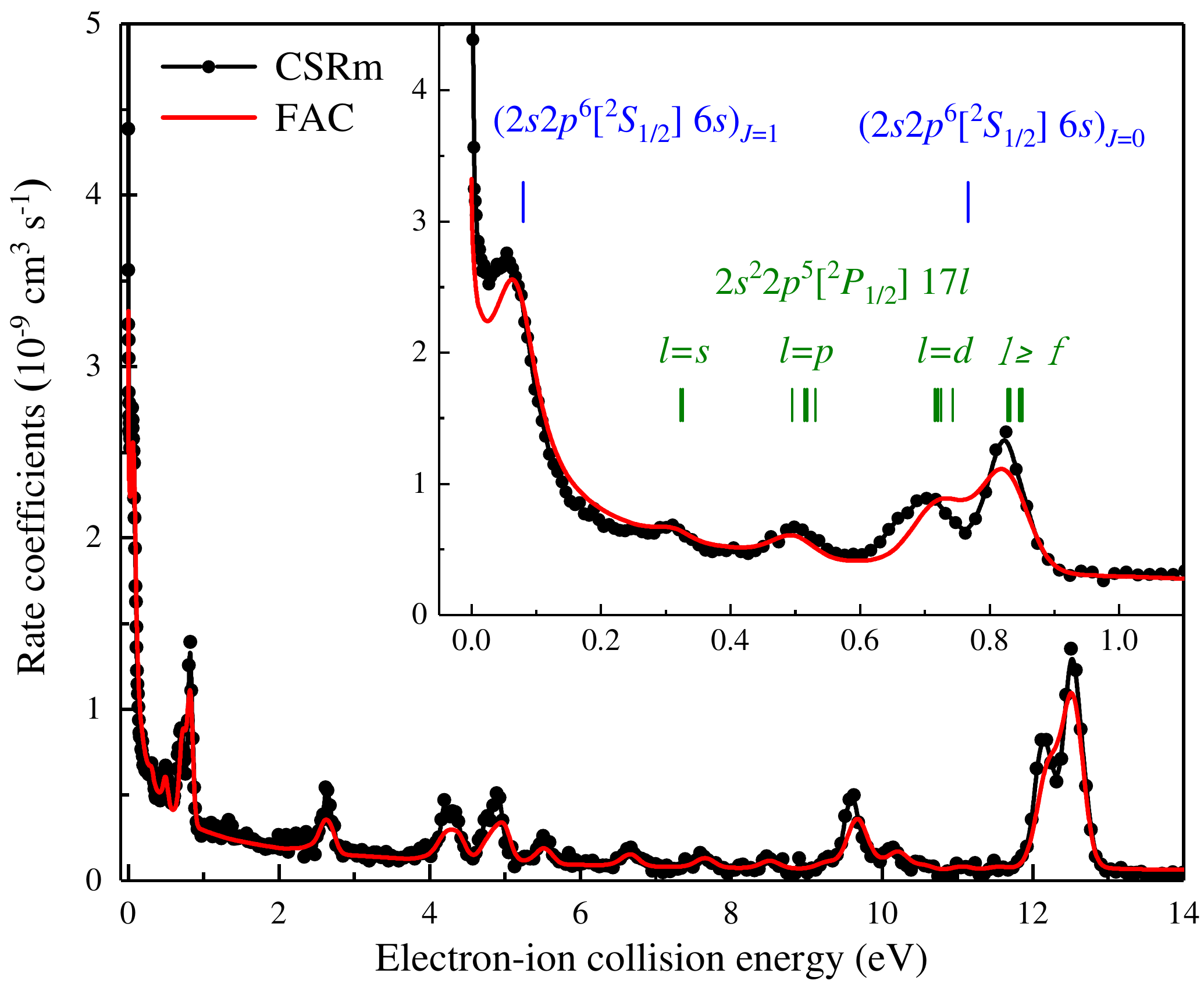}
	\caption{\label{fig:exp}Experimental electron-ion recombination spectrum (black symbols) in comparison with the result of  theoretical calculations utilizing the FAC code (red full line). The inset enlarges the low-energy resonances with their calculated resonance positions indicated by the vertical bars. }
\end{figure}

The recombined ion counts and the related parameters were stored each millisecond and the recorded data from the 
centers of the detuning periods were taken for deriving the rate coefficient in the data processing. The original 
spectra of the ion counts as a function of the detuning voltages up to $\pm$150~V are presented Fig.~\ref{fig:data}a.  
The data displayed in Fig.~\ref{fig:data}b resulted from transforming the detuning voltages into electron-ion  
collision energies via Eq.~(\ref{eq:E}) and the count rates into rate coefficients via Eq.~(\ref{eq:alpha}). It should 
be 
noted that in addition to DR resonances the displayed rate coefficient also contains a contribution by nonresonant 
radiative recombination (RR, discussed below), that decreases sharply with increasing $\vert E \vert$, as well as a 
practically constant background resulting from charge-changing collisions with residual-gas particles.

Figure~\ref{fig:exp} presents the background-subtracted experimental electron-ion recombination spectrum of 
$^{58}$Ni$^{19+}$ ions in the energy range of 0--14~eV together with a theoretical result that we obtained by using the 
FAC code \cite{Gu2008}. For the comparison the  calculated DR cross sections were convoluted with the experimental 
electron beam velocity distribution (Eq.~(\ref{eq:vdist})).  There is a good agreement with the measured data. Minor 
discrepancies are likely to be attributed to an approximate treatment of electron correlation effects  in the 
calculation. The inset magnifies the resonances below 1~eV, and the labeled vertical bars mark the calculated DR 
resonance positions. DR via  $2s\,2p^{6}[^{2}S_{1/2}]6s$ and $2s^{2}2p^{5}[^{2}P_{1/2}]17l$ intermediate levels are the 
dominant channels in this energy range. Due to the limited experimental resolution the associated fine-structure 
splittings are generally not fully resolved in the present measurement. However, an isolated resonance  occurs at about 
86~meV on top of the steeply decreasing RR rate coefficient. It is attributed to  the 
$(2s\,2p^{6}[^{2}S_{1/2}]6s)_{J=1}$ intermediate level.

\subsection{Uncertainty of the energy scale}

According to Eq.~(\ref{eq:E}), the accuracy of electron-ion collision energy depends on the  Lorentz factors of the 
electron and ion beams as well as on the angles between the two beams. The Lorentz factor of the ion beam is
\begin{equation}
	\gamma_i = 1 + \frac{E_i}{m_ic^{2}},
\end{equation}
where $E_i$ and $m_i$ are the ion kinetic energy and the ion mass, respectively. The energy equivalent of the latter 
can be calculated as $m_ic^2 \approx Am_uc^2 - 19m_ec^2$ which neglects the contributions from the electron binding 
energies. Using  $A= 57.9353424$ for  the atomic mass of the $^{58}$Ni atom \cite{Wang2012} and $m_ec^2/m_uc^2= 
5.48579909\times10^{-4}$ for the electron mass in atomic mass units \cite{Tiesinga2021} one arrives at $m_ic^2 \approx 
57.92492~m_uc^2$.

The Lorentz factor of the electron beam
can be calculated from the  cooler-cathode voltage $U_c$, the space-charge potential $U_s$ of the electron beam, and the detuning voltage $U_d$ as
\begin{equation}\label{eq:ge}
	\gamma_e = 1 + \frac{e(U_c+U_s+U_d)}{m_ec^2}.
\end{equation}
At cooling, electrons and ions move with the same velocity such that $\gamma_i=\gamma_e$ for $U_d=0$. The ion energy can be thus expressed as
$E_i = e(U_c+U_s)m_i/m_e$ and the Lorentz factor of the ion beam can be written as
\begin{equation}\label{eq:gi}
	\gamma_i = 1 + \frac{e(U_c+U_s)}{m_ec^2}
\end{equation}
Neglecting uncertainties of the particle masses, the uncertainty  $\Delta E_\mathrm{rel}$ of the electron-ion collision 
energy (Eq.~(\ref{eq:E})) thus depends on the uncertainty $\Delta E_d$ of the detuning energy $E_d =eU_d$, the 
uncertainty $\Delta E_0$ of the cooling energy $E_0=e(U_c+U_s)$, and the uncertainty $\Delta\theta$ of the 
angle~$\theta$, i.e.
\begin{eqnarray}
\lefteqn{(\Delta E_\mathrm{rel})^2 =  \left\vert\frac{\partial E_\mathrm{rel}}{\partial \theta}\right\vert^2(\Delta\theta)^2\;+}\nonumber \\
&  & \left\vert\frac{\partial E_\mathrm{rel}}{\partial \gamma_i}\right\vert^2\left(\frac{\Delta E_0}{m_ec^2}\right)^2 
+\; \left\vert\frac{\partial E_\mathrm{rel}}{\partial \varepsilon_d}\right\vert^2\left(\frac{\Delta 
E_d}{m_ec^2}\right)^2 \label{eq:Delta}
\end{eqnarray}
The partial derivatives in Eq.~(\ref{eq:Delta}) can be straight-forwardly calculated after substituting $\gamma_e = 
\gamma_i+\varepsilon_d$ in Eq.~(\ref{eq:G}) with $\varepsilon_d =E_d/(m_ec^2)$.

The cooling energy can be obtained from the cathode voltage and the space charge potential. Both values bear 
considerable uncertainties. The cathode voltage can be read from the power supply with an uncertainty of 0.5\%. The 
space charge potential and its uncertainty are even less accessible since $U_s$  cannot be measured directly, although 
its influence on the ion-beam can be monitored by the Schottky beam analysis. In this situation we determined the ion 
energy and, thus, the cooling energy $E_0=E_im_e/m_i$ from the magnetic rigidity of the storage ring's dipole magnets, 
which keep the ions on a closed orbit. The associated uncertainty results in a relative uncertainty $\Delta E_0/E_0 = 
0.005$.  We also point out, that an accurate value of the cooling energy is required for obtaining recombination 
spectra that are symmetric about $E_\mathrm{rel}=0$ on the electron-ion collision-energy scale (Fig.~\ref{fig:data}).

\begin{figure}[t]
	\centering
	\includegraphics[width=1\linewidth]{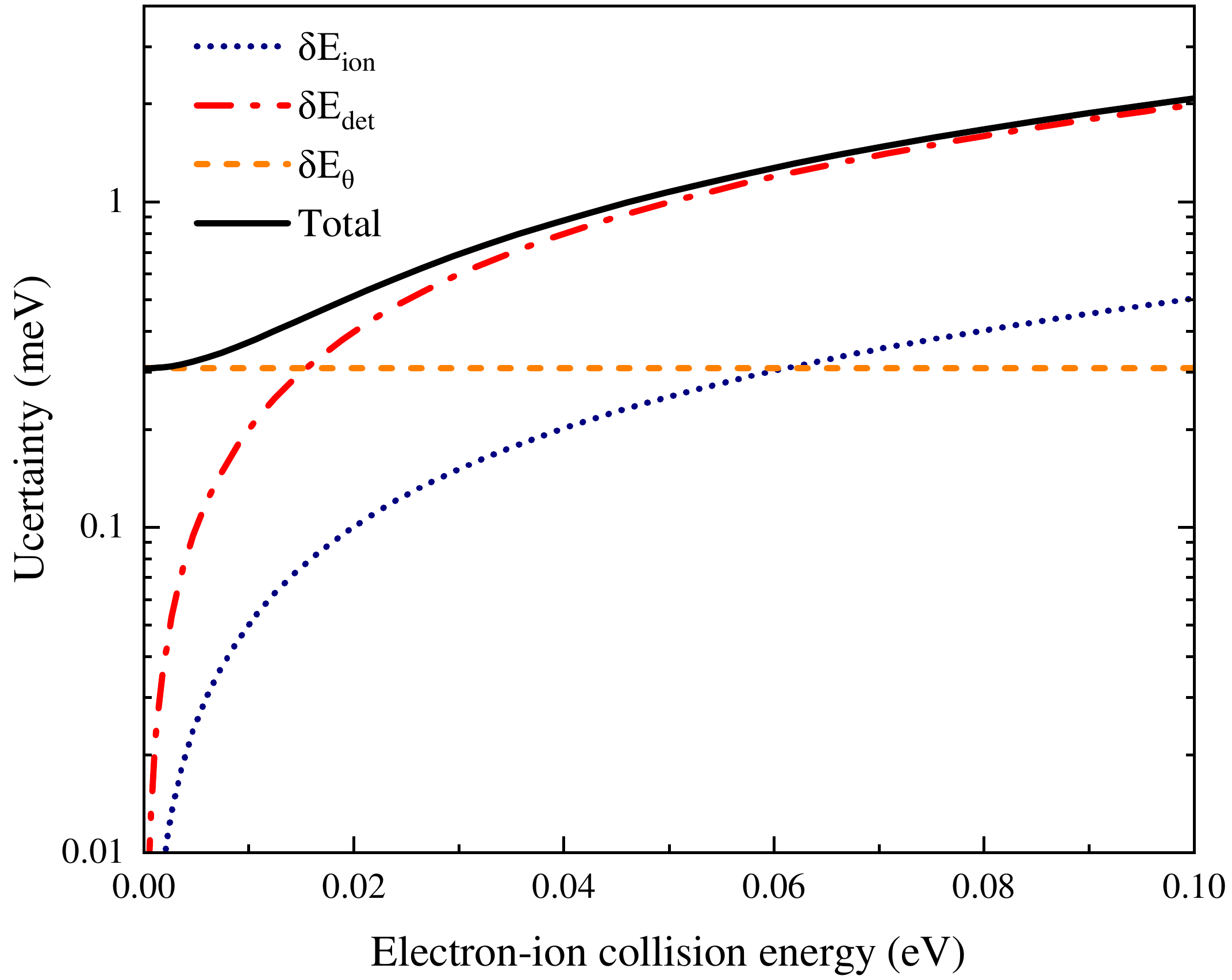}
	\caption{Systematic uncertainties of the electron-ion collision-energy scale. The individual contributions are 
	summed quadratically to obtain the total uncertainty (Eq.~(\ref{eq:Delta})). At the position of the lowest-energy 
	DR	resonance at 86~meV is amounts to 2.0~meV (Tab.~\ref{tab:err}).}
	\label{fig:err}
\end{figure}

The known Rydberg series limit of the $2s\,2p^{6}[^{2}S_{1/2}]nl$ DR resonances at $149.05\pm0.12$~eV 
\cite{Kramida2021} provides another calibration point of the energy scale. We applied a 2\% correction to the nominal 
collision-energy scale  to achieve agreement of  the experimental series limit with the known value \cite{Wang2019}. 
The 0.08\% relative uncertainty of the known series limit is a lower limit for the uncertainty of the thereby 
calibrated detuning energies. This calibration to a certain extent also takes care of ion-beam dragging effects that 
occur at non-zero relative energies, i.e, when $U_d\neq0$. The dragging effect is caused by the Coulomb interaction 
between the electron and ion beams which results in a force on the ion beam, trying to pull the ions to the current 
electron beam velocity \cite{Shi2002a,Kieslich2004a}. It should be noted, that the presently applied measurement mode 
(Fig.~\ref{fig:ramp})  with alternatingly negative and positive detuning energies applied for only short time intervals 
and considerably longer intervening cooling time intervals was deliberately chosen to minimize the drag-force effect. 
During the measurement, the Schottky spectrum also showed no evident signals of dragging. In order to account for 
possible nonlinearities of the $U_d$ power supply and the drag-force effect we conservatively assume a 1\% relative 
uncertainty for the detuning energy, i.e., $\Delta E_d/E_d = 0.01$.

The angle $\theta$ between the ion beam and the electron beam has been adjusted to 0 as best as achievable by using 
correction coils that are available at the cooler. The remaining uncertainty is estimated to be 0.3~mrad. Since 
$\partial E_\mathrm{rel}/\partial\theta \propto \sin\theta$ one obtains a zero error for $\theta=0$ from this 
derivative. Therefore we have calculated the contribution of $\Delta\theta$ to the error budget as the difference 
$E_\mathrm{rel}(\gamma_i,\varepsilon_d,\theta=0.3\mathrm{~mrad}) - E_\mathrm{rel}(\gamma_i,\varepsilon_d,\theta=0)$.

Figure~\ref{fig:err} visualizes the individual contributions to the total uncertainty which correspond to the summands 
on the right-hand side of Eq.~(\ref{eq:Delta}). Numerical values for these uncertainties at the position of the 
lowest-energy DR resonance are given in Tab.~\ref{tab:err}. The total systematic uncertainty of the resonance position 
is $\pm$2~meV. Additional statistical and fitting uncertainties will be discussed below.

\begin{table}
	\caption{\label{tab:err}Systematic experimental uncertainties of the electron-ion collision energy 86~meV, i.e., at the position of the lowest-energy DR-resonance (Fig.~\ref{fig:exp})}
	\begin{ruledtabular}
		\begin{tabular}{lc}	
			& $\Delta E_\mathrm{rel}$ (meV)   \\
			\hline
			Angle between beams      & 0.31 \\
			Detuning voltage      & 1.69 \\
			Ion beam energy       & 0.43 \\
			Energy calibration    & 0.86 \\
\hline
			Total (quadratic sum) & 1.97
		\end{tabular}
	\end{ruledtabular}
\end{table}

\section{Theoretical treatment}\label{sec:theory}
\subsection{MCDHF calculation}
The $2s^{2}2p^{5}[^{2}P_{3/2}] \to 2s2p^{6}[^{2}S_{1/2}]$ transition energy and the binding energy of the 6$s$ electron in the $2s2p^{6}[^{2}S_{1/2}]6s$ state are calculated by using the relativistic atomic structure package GRASP2018 \cite{FroeseFischer2019}, which was developed based on the multi-configurational Dirac-Hartree-Fock (MCDHF) method \cite{Grant2007, Desclaux1975}. In this method, an atomic-state wave
function with a specific parity $P$, total angular momentum $J$, and its projection $M$ on the quantization axis is approximated by a set of configuration-state wavefunctions (CSFs) with the same $PJM$ as follows \cite{Grant2007, Desclaux1975},
\begin{eqnarray}
\label{eq9-wave function}
\psi_{\alpha}(PJM)=\sum_{r=1}^{n_{c}}c_{r}(\alpha)|\phi_{r}(PJM)\rangle \, .
\end{eqnarray}
Here, $n_{c}$ is the number of the CSFs used. ${c_{r}(\alpha)}$ denotes configuration mixing coefficients, which give rise to a representation of the atomic state $|\psi_{\alpha}\rangle$ in the chosen basis $\{|\phi_{r}\rangle\}$. The CSFs are initially generated as an anti-symmetrized product of a set of orthonormal orbitals and, then, optimized self-consistently in the basis of the Dirac-Coulomb-Breit Hamiltonian, which is followed by an inclusion of the quantum-electrodynamical effects into the representation ${c_{r}(\alpha)}$ of the atomic state $|\psi_{\alpha}\rangle$ by diagonalizing the Dirac-Coulomb-Breit Hamiltonian matrix. In the calculations of the excitation energy from $2s^{2}2p^{5} \, [^{2}P_{3/2}]$ to $2s2p^{6} \, [^{2}S_{1/2}]$ in fluorine-like Ni$^{19+}$ ion, 
all the single and double substitutions from the multi-reference (MR) configurations (the $ 2s^2 2p^5 $, $2s 2p^6 $, $2s^2 2p^4 3p$, and $2s 2p^5 3p$)  to the active set \{$9s$, $9p$, $9d$, $9f$, $9g$, $9h$, $9i$, $9k$\} are considered, which generate 686327 CSFs for the block of $J=3/2$ and odd parity,  and 274090 CSFs for the block of $J=1/2$ and even parity, respectively.
Moreover, in obtaining the energy level $(2s2p^{6}[^{2}S_{1/2}]6s)_{J=1}$ of neon-like Ni$^{18+}$ ion, all the single and double substitutions from the MR configurations (the $2s 2p^6 5s$, $2s 2p^6 6s$,  and $2s 2p^6 7s$)  to the active set \{$13s$, $13p$, $13d$, $13f$, $13g$, $12h$, $11i$, $10k$\} are considered, which generate 522644 CSFs for the block of $J=1$ and even parity.

\begin{table*}[!htbp]

		\caption{\label{Table_contribution}
	Total energies ($E_h$) of  the $2s^{2}2p^{5} \, [^{2}P_{3/2}]$ and $2s2p^{6} \, [^{2}S_{1/2}]$  levels of 
	fluorine-like Ni$^{19+}$ ion and the $(2s2p^{6}[^{2}S_{1/2}]6s)_{J=1}$ level  of neon-like Ni$^{18+}$ ion are 
	presented as a function of the increasing active set (AS), as well as the transition energies $\Delta E$ (eV)  for 
	the transitions  $2s^22p^5\;^2P_{3/2}\to 
	2s\,2p^6\;^2S_{1/2}$
	 and $(2s2p^{6}[^{2}S_{1/2}]6s)_{J=1} \to 2s2p^{6} \, [^{2}S_{1/2}]$.
		}

	\begin{tabular}{lccc}

	\colrule 	\colrule
	\multicolumn{4}{c}{F-like Ni$^{19+}$ (MR = \{$2s^2 2p^5 $, $2s 2p^6 $, $2s^2 2p^4 3p$, $2s 2p^5 3p$\})}   \\
	\colrule
 \multicolumn{1}{c}{AO} & \multicolumn{1}{c}{$E~(2s^{2}2p^{5} \, [^{2}P_{3/2}])$} & \multicolumn{1}{c}{$E~(2s2p^{6} \, [^{2}S_{1/2}])$} & \multicolumn{1}{c}{$\Delta E $($2s2p^{6} \, [^{2}S_{1/2}] \to 2s^{2}2p^{5} \, [^{2}P_{3/2}]$)} \\
					\colrule			
 \{$3s$, $3p$, $3d$\}                                      & -1292.5833877 & -1287.0299079 & 151.118 \\
 \{$4s$, $4p$, $4d$, $4f$\}                                & -1292.7902178 & -1287.3083550 & 149.169 \\
 \{$5s$, $5p$, $5d$, $5f$, $5g$\}                          & -1292.8534402 & -1287.3745748 & 149.088 \\
 \{$6s$, $6p$, $6d$, $6f$, $6g$, $6h$\}                    & -1292.8806642 & -1287.4035955 & 149.039 \\
 \{$7s$, $7p$, $7d$, $7f$, $7g$, $7h$, $7i$\}              & -1292.9096358 & -1287.4331842 & 149.022 \\
 \{$8s$, $8p$, $8d$, $8f$, $8g$, $8h$, $8i$, $8k$\}        & -1292.9238447 & -1287.4474401 & 149.020 \\
 \{$9s$, $9p$, $9d$, $9f$, $9g$, $9h$, $9i$, $9k$\}        & -1292.9295149 & -1287.4531798 & 149.019 \\
\colrule \colrule
\multicolumn{4}{c}{Ne-like Ni$^{18+}$ (MR = \{$2s 2p^6 5s$, $2s 2p^6 6s$,  $2s 2p^6 7s$\})}\\
	\colrule
\multicolumn{1}{c}{AO} && \multicolumn{1}{c}{$E ((2s2p^{6}[^{2}S_{1/2}]6s)_{J=1})$} &\multicolumn{1}{c}{$\Delta E $($(2s2p^{6}[^{2}S_{1/2}]6s)_{J=1} \to 2s2p^{6} \, [^{2}S_{1/2}]$) } \\
		\colrule
\{$8s$, $8p$, $8d$, $8f$, $8g$, $8h$, $8i$, $8k$\}         &               & -1292.8391704 & 158.08  \\
\{$9s$, $9p$, $9d$, $9f$, $9g$, $9h$, $9i$, $9k$\}         &               & -1292.8712688 & 151.37  \\
\{$10s$, $10p$, $10d$, $10f$, $10g$, $10h$, $10i$, $10k$\} &               & -1292.8918299 & 150.13  \\
\{$11s$, $11p$, $11d$, $11f$, $11g$, $11h$, $11i$, $10k$\} &               & -1292.9022762 & 149.63  \\
\{$12s$, $12p$, $12d$, $12f$, $12g$, $12h$, $11i$, $10k$\} &               & -1292.9085850 & 148.99  \\
\{$13s$, $13p$, $13d$, $13f$, $13g$, $12h$, $11i$, $10k$\} &               & -1292.9165405 & 148.82  \\
\{$14s$, $14p$, $14d$, $14f$, $13g$, $12h$, $11i$, $10k$\} &               & -1292.9277362 & 148.97  \\		
	\colrule
	\end{tabular}

\end{table*}

In Table~\ref{Table_contribution}, total energies ($E_h$) of  the $2s^{2}2p^{5} \, [^{2}P_{3/2}]$ and $2s2p^{6} \, 
[^{2}S_{1/2}]$  levels of fluorine-like Ni$^{19+}$ ion and the $(2s2p^{6}[^{2}S_{1/2}]6s)_{J=1}$ level  of neon-like 
Ni$^{18+}$ ion are presented as a function of the increasing active set (AS), as well as the transition energies 
$\Delta E$ (eV)  for the transitions  $2s^22p^5\;^2P_{3/2}\to 
2s\,2p^6\;^2S_{1/2}$ and $(2s2p^{6}[^{2}S_{1/2}]6s)_{J=1} \to 2s2p^{6} \, [^{2}S_{1/2}]$. The MCDHF calculated total 
energies and $\Delta E$ are well converged with respect to the increasing size of the AS.

\subsection{\emph{Ab-initio} calculation}

Different theoretical approaches, e.g., complex coordinate rotation method \cite{Ho1983}, Feshbach projection operator method, optical potential method, $R$-matrix method, etc. (see, e.g., Ref. \cite{Schneider2016} and references therein) have been proposed to describe the autoionizing states. In the present work, we apply the stabilization method (SM), pioneered by Hazi and co-workers \cite{Hazi1970,Fels1971} utilized in numerous investigations \cite{Mueller1994b, Kar2005, Saha2009, Amaro2021}. The idea of this method \cite{Mandelshtam1993} is to diagonalize the Hamiltonian of a quantum system with suitable square-integrable real wave functions and investigate the spectra in the neighborhood of resonance position under small variations of the basis set. It can be done elegantly using the spectral density of states function
\begin{equation}
\rho_n(E) = \left| \frac{\xi_{i+1} - \xi_{i}} {E_n(\xi_{i+1}) - E_n(\xi_{i})} \right|\,,
\end{equation}
where $\xi_i$ is the basis variation parameter, $E_n(\xi_i)$ is the energy level near the resonance position. The maximum of the spectra density function $\rho_n$ corresponds to the energy of resonance state (see, e.g., Ref. \cite{Amaro2021} for details). Our realization of the approach is based on the configuration-interaction Dirac-Fock-Strum (CI-DFS) method \cite{Tupitsyn2003, Tupitsyn2005, Kaygorodov2019}, where the basis set is varied by the reference energy parameter for Sturm basis orbitals.

The QED calculations of the transition $2s 2p^6\,^2S_{1/2} \to 2s^2 2p^5\,^2P_{3/2}$ energy in fluorine-like nickel is 
based on the QED perturbation theory in the extended Furry picture \cite{Furry1951}, which previously was also employed 
for the evaluation of the ground state fine structure energy in fluorine-like ions \cite{Volotka2019, Shabaev2020, 
ONeil2020, Lu2020c}. The zeroth-order Hamiltonian is defined as
\begin{equation}
	\label{eq:H0}
	H_0 = \int d^3x \psi^\dagger(x) \left[ -i{\bm \alpha\nabla} + \beta m + V_{\rm C}({\bf x}) + V_{\rm scr}({\bf x}) \right] \psi(x)\,,
\end{equation}
where $\alpha_i$ and $\beta$ are the Dirac matrices, $\psi(x)$ is a field operator expanded in terms of the Dirac wave functions $\psi_n({\bf x})$:
\begin{equation}
	\left[ -i{\bm \alpha\nabla} + \beta m + V_{\rm C}({\bf x}) + V_{\rm scr}({\bf x}) \right] \psi_n({\bf x}) = \varepsilon_n \psi_n({\bf x})\,.
\end{equation}
In Eq. (\ref{eq:H0}), in addition to the nuclear Coulomb potential $V_{\rm C}$ the screening potential $V_{\rm scr}$ which partially accounts for the interelectronic interaction has been added. In our calculations, we employ the core-Hartree and Kohn-Sham potentials.

The perturbation expansion is performed with respect to the interaction Hamiltonian
\begin{equation}
	\label{eq:HI}
	H_{\rm int} = \int d^3x \left[ \bar{\psi}(x) e \gamma^\mu A_\mu(x) \psi(x) - \psi^\dagger(x) V_{\rm scr}({\bf x}) \psi(x) \right]\,,
\end{equation}
where $\gamma^\mu = (\beta, \beta\alpha_i)$, $A_\mu(x)$ is a photon field operator. The second term in Eq. (\ref{eq:HI}) corresponds to the subtraction of the counter term. For constructing the perturbative expansion, we use the two-time Green function method \cite{Shabaev2002b}. In the present calculations, we account entirely for the first-order corrections, which are given by the one-photon exchange and the one-electron self-energy and vacuum polarization diagrams. These diagrams are computed employing the well-known formal expressions, which can be found in, e.g., Refs. \cite{Mohr1998, Shabaev2002b}. The second-order diagrams include the many-electron radiative (so-called screened QED), one-electron two-loop, and two-photon exchange corrections. Here, we evaluate only the screened QED correction employing the techniques and methods thoroughly presented in Refs. \cite{Kozhedub2010, Malyshev2014, Malyshev2015, Volotka2019}. For the one-electron two-loop contribution, we use the hydrogenic values from Ref. \cite{Yerokhin2015}. While the two-photon exchange diagrams are taken into account within the Breit approximation, and the uncertainty from missing higher-order correction is estimated to be $\alpha/(8\pi)(\alpha Z)^4/Z^2$ multiplied by a factor $5$.

The terms which were not accounted for by the rigorous QED theory have been evaluated within the Breit approximation 
employing the CI-DFS method. Within this method, the many-electron wave-function and the energy of an atom $E$ in the Breit 
approximation are to be found as solutions of the Dirac-Coulomb-Breit Hamiltonian. Thus, the CI-DFS method is used to 
calculate the second and higher orders interelectronic-interaction corrections. Moreover, we have evaluated the recoil 
correction employing the many-body relativistic mass shift Hamiltonian \cite{Shabaev1985,shabaev1988nuclear,Palmer1987, Shabaev1994, Kozhedub2010}.

\section{Results and discussion}\label{sec:results}

\begin{figure}[htbp]
	\centering
	\includegraphics[width=1\linewidth]{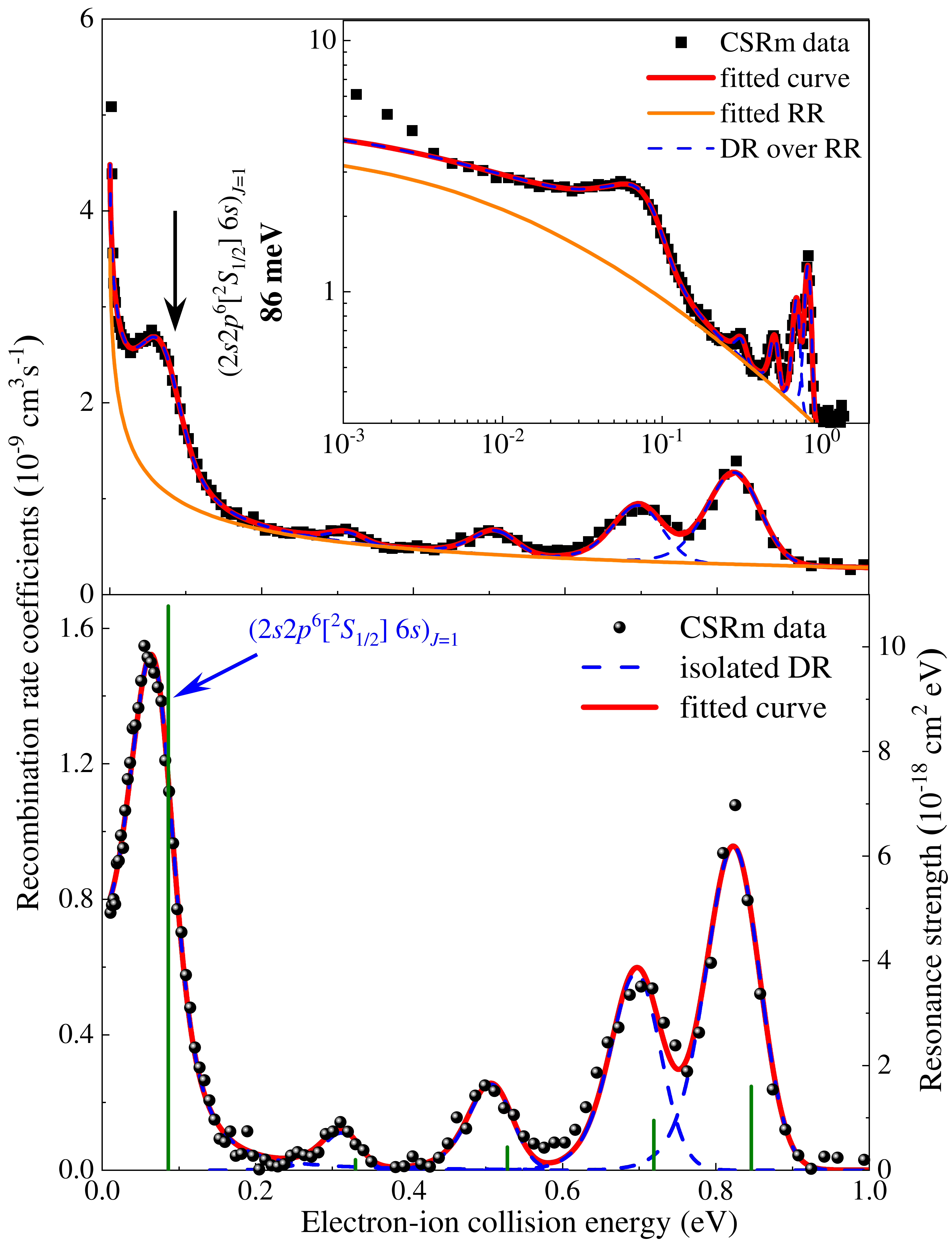}
	\caption{\label{fig:fit}(a) Fit (red solid line) of  the experimental low-energy DR spectrum (black symbols). The 
	fit accounts for isolated DR resonances (blue dashed lines) and the  RR rate coefficient  (orange solid line) . (b) 
	Experimental DR rate coefficient (symbols) after subtraction of a constant background and the fitted RR 
	contribution  along with the fit curve for the DR resonances. The vertical green bars represent the fitted DR 
	resonance positions and strengths.}
\end{figure}

The fitted curve of the low-energy DR spectrum is presented in Fig.~\ref{fig:fit}a. The data points below 10~meV were 
excluded from the fitting, due to the steep rise of the measured rate coefficient  towards lower energies (see 
Ref.~\cite{Wang2019} for details), which is difficult to account for in the fit.  In the fitting procedure, the first 
resonance was treated as a Lorentzian profile multiplied by a factor of $E_\mathrm{res}/E_\mathrm{rel}$ as discussed in 
Ref.~\cite{Schippers2004c}, while the remaining resonances are treated as delta functions. Moreover, the DR and RR 
model cross-sections were convolved with the electron beam velocity distribution (Eq.~(\ref{eq:vdist})) to account for 
the experimental energy spread. The electron-beam temperatures obtained from the fit are $k_{B}T_{\parallel} = 
0.56\pm0.05$ ~meV and $k_{B}T_{\perp} = 23\pm1$~meV.

According to the fit, the position of the lowest-energy resonance was determined to $85.8 \pm 1.2$~meV.  Its combined 
systematic and statistical (resulting from the fit)  uncertainty  is less than 4~meV at a one-sigma confidence level. 
The fitted resonance positions and strengths are  in fair agreement with the FAC calculation (Tab.~\ref{tab:res}) which 
assigns the  $(2s2p^{6}[^{2}S_{1/2}]6s)_{J=1}$  intermediate level to the lowest-energy resonance. In 
Fig.~\ref{fig:fit}b , the DR fit results are compared with the background subtracted experimental data to provide a 
better view of the DR resonances. The fact that the maxima of the resonances do not coincide with the fitted resonance 
positions is a consequence of the  asymmetry of the electron-velocity distribution (Eq.~(\ref{eq:vdist})).

\begin{table}
	\caption{\label{tab:res}Experimental and theoretical positions, natural widths and strengths for the lowest-energy 
	resonance as assigned to the $(2s2p^{6}[^{2}S_{1/2}]6s)_{J=1}$  intermediate level. The number in the brackets 
	indicate the experimental uncertainties.}
	\begin{ruledtabular}
		\begin{tabular}{lcc}	
			& CSRm  & FAC   \\
			\hline
			Resonance position (meV)   & 86(4)    & 80     \\
			Natural width (meV)        & 51(5)    & 56     \\
			Resonance strength (10$^{-18}$ eV$\cdot$cm$^{2}$)   & 10.8(1.0)  & 10.2  \\
		\end{tabular}
	\end{ruledtabular}
\end{table}

As described in Sec.~\ref{sec:theory}, we have calculated  the binding energy of the $6s$ electron of the 
$(2s2p^{6}[^{2}S_{1/2}]6s)_{J=1}$ level using the MCDHF and stabilization methods. We find that the large-scale MCDHF 
calculations do not converge on the binding energy as the active sets are increased. The energy of the autoionizing 
level turned out to be extremely sensitive to the choice of basis sets  because of the near-degeneracy with the 
continuum. Therefore, our MCDHF value for the $6s$ binding energy of 148.970~eV bears a rather large uncertainty which 
is estimated to be $\pm$20~meV. The stabilization method exhibits a more favorable convergence behavior. It yields a 
value of $148.946\pm0.006$~eV for the binding energy of the $6s$ electron. The values for the $2s^22p^5\;^2P_{3/2}\to 
2s\,2p^6\;^2S_{1/2}$ transition energy in fluorine-like nickel that results from adding  the calculated 6$s$ binding 
energies in the $(2s2p^{6}[^{2}S_{1/2}]6s)_{J=1}$ level and the experimentally derived DR resonance position are listed 
in Tab.~\ref{tab:final}.

\begin{table}[t]
	\caption{Comparison of the experimental and theoretical results for the $2s^22p^5\;^2P_{3/2}\to 
		2s\,2p^6\;^2S_{1/2}$ transition energy in fluorine-like nickel ion (in eV). Figures in parentheses represent 
		one-sigma uncertainties.
		\label{tab:final}}
	\begin{ruledtabular}
		\begin{tabular}{l l}
			Method & Energy (eV)                                           \\
			\hline
			\multicolumn{2}{c}{Experiment + theory}                        \\
			Exp. + MCDHF   & $149.056(4)_\mathrm{exp}(20)_\mathrm{theo}$   \\
			Exp. + SM          & $149.032(4)_\mathrm{exp}(6)_\mathrm{theo}$    \\
			\multicolumn{2}{c}{Theory}                                     \\
			MCDHF              & 149.019(10)                               \\
			$Ab~initio$        & 149.046(7)
		\end{tabular}
	\end{ruledtabular}
\end{table}

Table~\ref{tab:final} also list the results for the  $2s^22p^5\;^2P_{3/2}\to 
2s\,2p^6\;^2S_{1/2}$ transition 
energies of our fully relativistic MCDHF and \emph{ab-initio} QED calculations.  For this quantity, the large-scale 
MCDHF calculation yield a convergent value with an uncertainty of $\pm$20~meV when increasing the size of active sets.  
In the $ab~initio$ QED calculation, the zeroth-order Dirac result is extended by the correlation corrections evaluated 
within the Breit approximation, by the first- and second-order QED contributions, as well as by the recoil term. 
Calculations were performed employing two different starting potentials, core-Hartree and Kohn-Sham. The final results 
as presented in Table \ref{tab:theo} appear to be independent of the initial potentials. In Table \ref{tab:theo}, we 
present the individual theoretical contributions to the transition $2s^22p^5\;^2P_{3/2}\to 
2s\,2p^6\;^2S_{1/2}$ energy in fluorine-like nickel calculated as has been explained above in both utilized screening 
potentials. As one can see from the table the total results in both potentials perfectly agree with each other. The 
final uncertainty is dominated by the estimation of the QED effect for the two-photon exchange correction.

\begin{table}[!htbp]
\caption{Individual contributions to the transition $2s^22p^5\;^2P_{3/2}\to 
	2s\,2p^6\;^2S_{1/2}$ energy in fluorine-like nickel ion (in eV).
	\label{tab:theo}}
\begin{tabular}{lSS}
	\hline\hline
	{Contribution}  &   {Core-Hartree} & {Kohn-Sham}  \\ \hline
	Dirac           & 123.911      & 128.743    \\
	Correlation (1) &  27.190      &  22.723    \\
	Correlation (2) & -1.536       & -1.972     \\
	Correlation (3) &   0.032(2)   &  0.102(2)  \\
	QED (1)         &  -0.506      & -0.510     \\
	QED (2)         &  -0.033(6)   & -0.028(6)  \\
	Recoil          &  -0.012(3)   & -0.012(3)  \\
	Total           & 149.046(7)   & 149.046(7) \\
	\hline\hline
\end{tabular}
\end{table}

\begin{figure}[t]
	\centering
	\includegraphics[width=1\linewidth]{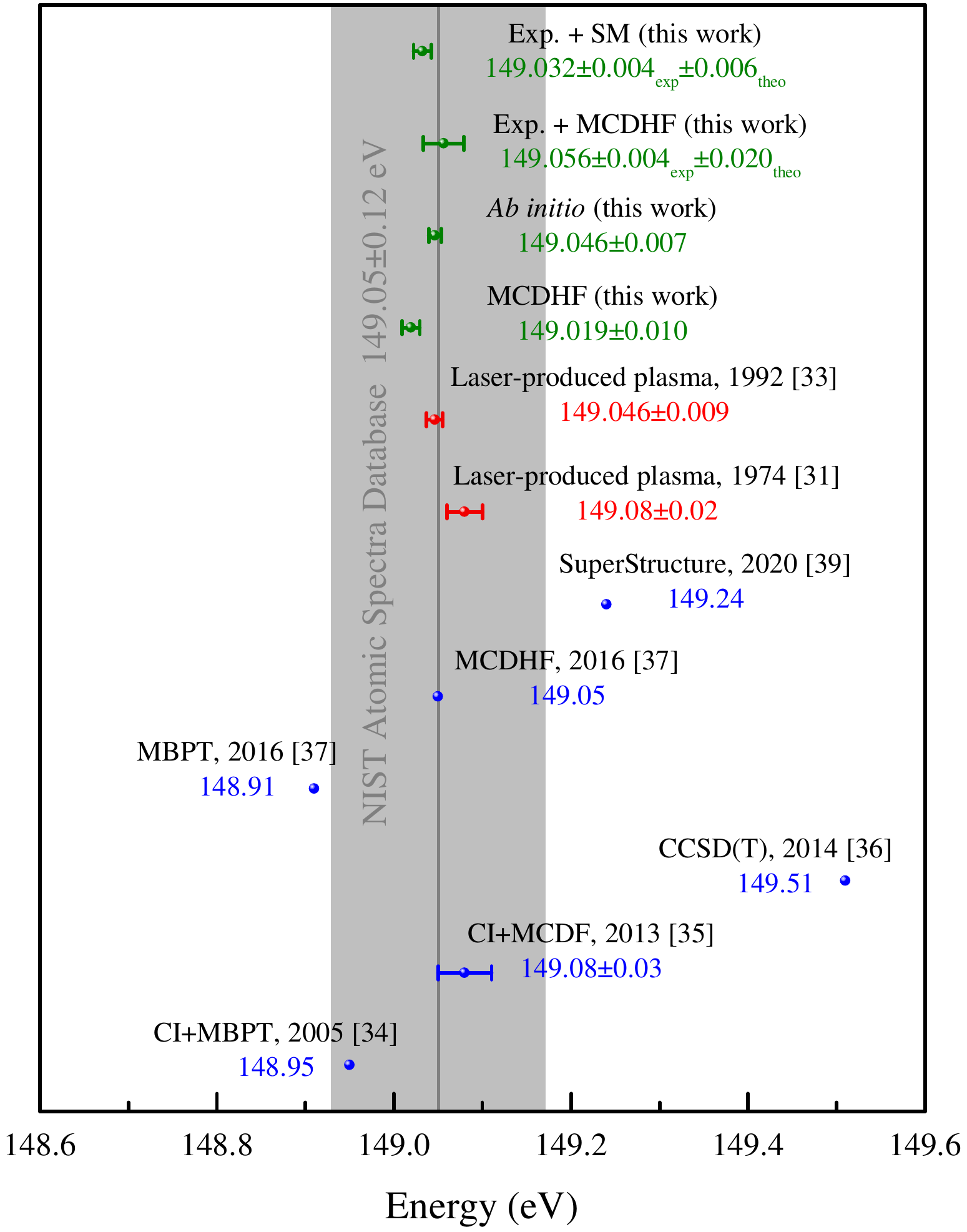}
	\caption{\label{fig:comp}Available experimental and theoretical transition energies of the 
	$2s^22p^5\;^2P_{3/2}\to 
	2s\,2p^6\;^2S_{1/2}$ along with the present experimentally-derived and fully theoretical results. The 
	vertical lines corresponds to the currently recommended value from the NIST Atomic Spectra Database 
	\cite{Kramida2021}. The grey shaded area marks the associated uncertainty.}	
\end{figure}

Figure \ref{fig:comp} compares the present experimentally-derived and theoretical results with the previous results 
from the literature. The determined transition energies agree with the most accurate plasma observation 
\cite{Sugar1992} within the error bars. The calculated values by the SuperStructure code \cite{Celik2020} and by the 
coupled cluster method with single and double excitations (CCSD) \cite{Nandy2014} are significantly larger than our 
results. The MBPT \cite{Gu2005a,Si2016} and MCDHF \cite{Si2016} calculations report values for the transition energies 
without uncertainties, what hampers an accurate comparison. The value obtained within the CI-MCDF method 
\cite{Joensson2013c} agrees well with both our experimental and theoretical results within the  given error bars, but 
our values are more precise. The individual contributions in Table \ref{tab:theo} indicate that third-order correlation 
effects contribute at least 0.032(2) eV to the total transition energy. The calculations where the correlation effect 
was handled with care \cite{Si2016,Joensson2013c,Gu2005a} yielded values which agree better with the experimental data 
as compared to the simpler approaches. The MBPT calculations \cite{Gu2005a,Si2016} are significantly lower than the 
present data indicating that the many-body expansion in many-electron systems remains a challenge task.

\section{Summary and outlook} \label{sec:sum}

Electron-ion recombination rate coefficients of fluorine-like nickel ions have been measured at the heavy-ion storage 
ring CSRm by employing the merged-beams method. The measured rate coefficients agree well with the most recent 
theoretical calculation by the FAC code, even at very low collision energies. The level-resolved theoretical 
calculation facilitates  the identification of the measured DR resonances. Accordingly, the lowest-energy  isolated 
resonance is associated with the $(2s2p^{6}[^{2}S_{1/2}]6s)_{J=1}$ intermediate state. Its experimental resonance 
position has been extracted by a fit to the measured recombination spectrum resulting in a value of  86~meV with a 
systemic uncertainty of 2 meV and a fitting error of 2 meV. This  accurate measurement of the resonant position  in 
combination with precision theoretical calculations of the binding energy of the $6s$ Rydberg electron enables a 
precise determination of the 	$2s^22p^5\;^2P_{3/2}\to 2s\,2p^6\;^2S_{1/2}$ core-transition energy. At the 
present level of experimental accuracy our results are sensitive to third-order correlation and second-order QED 
effects. The present study establishes  precision DR spectroscopy with highly charged ions at the CSRm and paves the 
way for future precision studies with highly-charged ions at the CSRe and the upcoming HIAF facility \cite{hiaf}.

\begin{acknowledgments}
This work has been funded by the National Key R\&D Program of China under Grant No. 2017YFA0402300; the National Natural Science Foundation of China through Grants No. U1932207, No. 11904371, No. 11674066, and No. 12104437, the Strategic Priority Research Program of Chinese Academy of Sciences Grant No. XDB34020000, and the Heavy Ion Research Facility in Lanzhou (HIRFL). S.X.W. is grateful to the Natural Science Foundation of Anhui Province (Grant No. 2108085QA27), and the Fundamental Research Funds for the Central Universities. W.Q.W. thanks the support from the Youth Innovation Promotion Association of the Chinese Academy of Sciences. A.V.V. acknowledges financial support by the Government of the Russian Federation through the ITMO Fellowship and Professorship Program. The authors would like to thank the CSR accelerator staff for their technical support during the experiment.
\end{acknowledgments}

\bibliography{Ni19DRrefs}
\end{document}